# MAGNETIC FIELD-ASSISTED GENE DELIVERY: ACHIEVEMENTS AND THERAPEUTIC POTENTIAL

José I. **Schwerdt**[1], Gerardo F. **Goya**[2,3], Pilar **Calatayud**[2], Claudia B. **Hereñú**[1], Paula C. **Reggiani**[1], Rodolfo G. **Goya**[1]

[1] *Institute for Biochemical Research-Histology B-Pathology B, Faculty of Medicine, National University of La Plata, La Plata, Argentina;*
[2]*Instituto de Nanociencia de Aragón (INA), University of Zaragoza, Zaragoza;*
[3]*Departamento de Física de la Materia Condensada, Facultad de Ciencias, University of Zaragoza, Zaragoza, Spain.*

## ABSTRACT

The discovery in the early 2000's that magnetic nanoparticles (MNPs) complexed to nonviral or viral vectors can, in the presence of an external magnetic field, greatly enhance gene transfer into cells has raised much interest. This technique, called magnetofection, was initially developed mainly to improve gene transfer in cell cultures, a simpler and more easily controllable scenario than in vivo models. These studies provided evidence for some unique capabilities of magnetofection. Progressively, the interest in magnetofection expanded to its application in animal models and led to the association of this technique with another technology, magnetic drug targeting (MDT). This combination offers the possibility to develop more efficient and less invasive gene therapy strategies for a number of major pathologies like cancer, neurodegeneration and myocardial infarction. The goal of MDT is to concentrate MNPs functionalized with therapeutic drugs, in target areas of the body by means of properly focused external magnetic fields. The availability of stable, nontoxic MNP-gene vector complexes now offers the opportunity to develop magnetic gene targeting (MGT), a variant of MDT in which the gene coding for a therapeutic molecule, rather than the molecule itself, is delivered to a therapeutic target area in the body. This article will first outline the principle of magnetofection, subsequently describing the properties of the magnetic fields and MNPs used in this technique. Next, it will review the results achieved by magnetofection in cell cultures. Last, the potential of MGT for implementing minimally invasive gene therapy will be discussed.

## INTRODUCTION

Gene therapy has undergone a remarkable development in the last 20 years. Particularly important advances have been made in the improvement of gene transfer and expression technology, with current efforts focusing on the design of safer and longer-expression gene vectors as well as systems possessing cell-type specificity for transgene delivery and regulatability of its expression by small molecules.

The association of viral vector-based gene delivery with nanotechnology now offers the possibility to develop more efficient and less invasive gene therapy strategies for a number of major pathologies including, but not limited to, cancer, neurodegeneration and myocardial infarction. This approach combines Magnetic Drug Targeting (MDT) and magnetofection, two methodologies based on the use of magnetic nanoparticles (MNPs). The concept of MDT was introduced by Widder *et al.* **[1]** and its goal was to concentrate magnetically responsive therapeutic complexes in target areas of the body by means of external magnetic fields. So far, the main application of MDT has been cancer therapy. Typically, magnetic microparticles (µm sized) or MNPs (nm sized) associated to a therapeutic drug are intravascularly injected near the tumor blood supply and are concentrated into the tumor by means of an external magnetic field. This strategy has shown promising results in clinical trials **[2, 3, also see below].** Magnetofection is a methodology developed in the early 2000's **[4, also see below]**. It is based on the association of MNPs with nonviral or viral vectors in order to optimize gene delivery in the presence of a magnetic field. The availability of stable, nontoxic MNP-gene vector complexes now offers the opportunity to implement magnetic gene targeting (MGT) in suitable animal models. MGT represents a variant of MDT in which the gene coding for a therapeutic molecule, rather than the molecule itself, is delivered to a therapeutic target area in the body. The advantage of MGT over MDT lies in the fact that in the former, when a vector complex unit transduces a target cell it generates large numbers of therapeutic molecules (amplification effect) for an extended period of time. If these are secreted molecules they will be released into the intercellular space.

This article will first outline the principle of magnetofection, subsequently describing the properties of the magnetic fields and MNPs used in this technique. Next, it will review the results achieved by magnetofection in cell cultures. Last, the potential of MGT for implementing minimally invasive gene therapy will be discussed. For a highly comprehensive review on magnetically-enhanced nucleic acid delivery the reader is referred to a recent article by Plank *e. al* **(5)**.

## MAGNETOFECTION

As indicated above, magnetofection is a methodology based on the association of MNPs with gene vectors in order to enhance gene transfer in the presence of a magnetic field. It was developed by Christian Plank and collaborators for gene transfer in cell cultures and *in vivo* using MNP-naked DNA complexes or MNP-viral vector complexes **[4].** In this context the principle of magnetofection in cells was assumed to be simple **(Fig. 1)**: the MNP-DNA complex is added to a culture of adherent cells and a magnet, placed close below the bottom of the flask or plate, attracts the magnetic complexes to the bottom where they come in close contact with the cells and are physically internalized, without any particular effect of the magnetic force on the endocytic uptake mechanism **[6].** For MNP-viral vector complexes it was thought that the magnetic field brought the complexes close to the cells thus favoring their internalization through viral receptor-mediated mechanisms. This results in a transduction improvement that in some cases is remarkable. For instance, in HEK293 cell cultures exposed to MNP-adenoviral vector complexes, magnetofection may induce over a 50-fold increase in transduction levels **(Fig. 2)**. For MNP-adenoviral vector complexes, the internalization mechanism outlined above does not hold, as suggested by the fact that certain cell lines (e.g., NIH3T3, K562 and primary human peripheral blood lymphocytes) which express little or no coxsackie virus and adenovirus (CAR) receptors and are therefore refractory to adenovectors, can be successfully transduced by magnetofection using MNP-adenovector complexes **[4].** Furthermore, ultrastructural analysis of MNP-recombinant adenovirus (RAd) complexes by electron and atomic force microscopy showed structurally intact adenoviruses fully surrounded by magnetic particles that occasionally bridged several virus particles **[7]**. Since this configuration would prevent virions from binding to their cell receptors, a still unknown internalization mechanism is likely to be involved. Also, kinetic studies with gold/iron oxide-based MNP-RAd complexes in adenovirus resistant cell lines provided additional evidence for a non receptor-mediated internalization mechanism for RAd-MNP complexes **[8].**

Regardless of the mechanisms by which magnetofection enhances gene transfer, over the years this technique has demonstrated to be highly effective in cell cultures (see below) and constitutes a promising tool for the implementation of MGT in vivo (see below). The commercial availability of magnetofection reagents has made this methodolgy readily accessible to nonspecialist researchers.

Besides a gene vector, two other key components are necessary to implement magnetofection namely, a suitable magnetic field applicator and properly formulated MNPs.

## PHYSICAL PROPERTIES OF MAGNETS AND MNP

For any biomedical application using MNP-based vectors, the magnetic component of these complexes (i.e., the magnetic core) needs to be specifically designed and engineered regarding its chemical and magnetic properties, so that the magnetic interaction can be maximized. The underlying physical interaction is related to the force generated on the magnetic core of any MNP-based complex, when a magnetic particle with magnetic (dipole) moment μ is placed in a non-uniform magnetic field **B**. In such a non-uniform magnetic field the force **F** exerted on a magnetic dipole with value (μ is related to the spatial variation, assumed in the x-direction) of **B** through its spatial derivatives

$$\boldsymbol{F} = (\boldsymbol{\mu} \cdot \boldsymbol{\nabla})\mathbf{B} \qquad \text{(equation 1)}$$

**Figure 3** shows the magnetic field **B** numerically simulated for a disk-shaped permanent magnet (NdFeSm) having 2 cm in diameter and 1 cm in height . The **B** profile was obtained applying a finite element method (FEM) on the corresponding Maxwell equations and boundary conditions. It can be seen that the field **B** decreases from a maximum value at the surface to a 25% of this value just 1 cm away from the magnet surface. Moreover the magnetic force, proportional to the derivative dB/dx, drops similarly within the same 1 cm distance, making it difficult to apply this simple method for obtaining constant forces within any practical working volume.

From equation (1) it is clear that, from a physical point of view, the MNPs must display the highest possible magnetic moment, which is related to the saturation magnetization $M_S$ (at room temperature) of the core materials. Compounds having large $M_S$ include pure 3d transition metals, which are extremely difficult to stabilize against oxidation in biological media (**Table 1)**. Among those oxides having large $M_S$, rare earths (Nd or Sm) or Ba are unsafe materials regarding toxicity levels. The iron oxides $Fe_3O_4$ (magnetite) and $\alpha$-$Fe_2O_3$ (hematite) are the only materials already approved for human applications in a variety of clinical protocols. Magnetite has been shown to fulfill the requirements of high Curie temperature ($T_C$), high saturation magnetic moment ($M_S$ ~ 90-98 emu/g, or ~450-500 emu/cm$^3$) and low toxicity. Although from the production point of view magnetite is cheap and relatively easy to obtain in highly purified form, the manufacture of MNPs featuring magnetically ordered cores of few nm in diameter is a major challenge because the high surface/volume ratio causes superficial disorder effects to become dominant.

In magnetofection, the magnetic field is applied to move the MNP-gene vector complexes towards the target site. In practice, this means that the target site ought to be subject to a magnetic flux density which is sufficient to cause saturation magnetization of the magnetic complex and ought to be subject to the highest possible field gradient. For magnetofection in cell cultures this requirement is not difficult to fulfil but for in vivo applications, magnets need to be tailor-made according to the anatomy of the target region in order to optimize magnetic trapping of complex particles. Magnets in the 96-well microtiter plate format are commercially available. In these plates Nd-Fe-B cylindrical magnets are assembled in antiparallel arrays. They produce a magnetic flux density ranging from 0.13 to 0.24 T. In contrast, most of the magnets used for in vivo studies have not been optimized in design and shape **(9).**

## CHEMICAL AND BIOLOGICAL PROPERTIES OF MNP

Applications of MNPs in biomedical areas require the use of a *colloidal ferrofluid*, or *magnetic colloids*, which consist of a suspension of magnetic particles of nanometric sizes in aqueous biological fluids (e.g., serum or cerebrospinal fluid (CSF)). These colloids usually have particle concentrations in the range of $10^{15}$-$10^{17}$ particles/ml. The stability of any magnetic colloid depends on the balance between attractive (van der Waals and dipole-dipole) and repulsive (steric and electrostatic) forces between the particles and the surrounding solvent molecules. Temperature is also a relevant parameter for stability due to energy transfer from the solvent molecules (Brownian motion) to the nanometric particles. Therefore, to stabilize the suspended MNPs against these forces they are often coated with a biocompatible polymeric layer. Nanoparticles stabilized by electrically neutral molecules (amphiphilic molecules, as oleic acid or alkylsilanes) constitute a *surfacted colloid.* Steric repulsion between particles acts as a physical barrier that keeps grains in suspension and stabilizes the colloid in nonpolar solvents. The polar heads of surfactant molecules can be cationic, anionic, zwitterionic or nonionic. A number of biocompatible surfactants/stabilizers have been used to generate MNPs. They include derivatized dextrans, starch or polycations such as polyethylenimine (PEI), polylysine (PL), protamine sulfate (PS), polycarylic acid and polybrene (PB, hexadimethrine bromide), among others **[10]**. In summary, the requirements that MNPs need to meet in order to be suitable for magnetofection are:

***Surface functionality.*** The surface of the coating layer of MNPs serves different purposes: (a) it stabilizes the MNPs in suspension and determines their shape during the growth process when they are produced; (b) it provides functional groups at the surface for further derivatization with organic groups or active biomolecules.

***Functional compatibility with the vector***. The association of MNPs with gene vectors or third components must not impair the functionality of the vectors concerning DNA delivery and expression.

***Biocompatibility.*** MNPs have to show low or negligible toxic effects on both cell cultures and in vivo. Different kinds of viability assays are to be performed before a given MNP is considered as non-toxic.

***Dispersion stability***. MNPs should be available as monodisperse (i.e., nonaggregated) particles suspended in suitable physiological fluids. Sample preparation should ensure stability against particle precipitation, aggregation and/or self assembly phenomena.

***High magnetic response***, in order to induce magnetic complex migration towards and concentration in the target area under the effect of an external magnetic field. Proper magnetic field profiles are also needed; they are usually designed by numerical simulation of magnet configurations. These calculations are in principle capable of engineering efficient magnetic field applicators particularly for in vivo use.

## SYNTHESIS OF MNPs

Magnetic nanoparticles can be produced by a number of physical and chemical routes that differ in the final properties of the products. For an overview on synthesis procedures and characteristics of nanoparticles suitable for gene delivery see **(11).** A broad classification scheme can be made based on the physical state of the starting materials. In the ***top-down*** strategy, the starting bulk material is reduced to nanometric scale in one (thin films), two (nanowires) or three (nanoparticles, or quantum dots) dimensions. This route is often based in physical processes like mechanical alloying, laser machining, laser chemical etching, reactive ion etching, among others. On the contrary, the ***bottom-up*** approach uses atomic or molecular units as starting materials to grow larger, nanometric structures. Bottom up techniques include chemical vapor deposition (CVD), reactive sputtering, plasma enhanced CVD, pulsed laser deposition (PLD), molecular beam epitaxy (MBE), and also wet routes like sol-gel and microemulsion thechniques. Most of the above techniques have attained good control of physical parameters of the products such as phase purity, particle shape, crystalline order and the attainable range of particle sizes, although tailoring all of these parameters in a single product remains a challenging task. Two main approaches for MNP synthesis can be considered.

**Thermal decomposition from organic precursors**

Monodisperse iron-oxide nanoparticles of different sizes ranging from 2 to 20 nm can be obtained by high temperature (250-350 ºC) decomposition of iron organic precursors **(Figure 4) [12].** These MNPs can be further functionalized with relevant biological molecules attached to the surface **[13]**. The synthesis method is based on the use of iron (or any transition metal) acetylacetonate (acac) and different solvents (e.g., phenyl ether or 1-Octadecene) which lead to different synthesis temperatures. To control the final particle size, different precursor/surfactant molar ratios can be used **[14]**.

This single-step, high-temperature synthesis for $Fe_3O_4$ MNPs is governed by the thermal decomposition of the precursor $Fe(acac)_3$ in the presence of a long-chain alcohol (e.g., 1,2 octanediol) and surfactants (oleic acid and oleilamine) using phenyl ether (boiling point ~533 K) as organic medium. For MNPs with $<d> < 10$ nm, this process can be further modified to tailor the final particle sizes through the molar ratio $[Fe(acac)_3]$/[surfactants] as reported by Vargas et al. **[15]**. The nanoparticles obtained usually range from 3 to 12 nm, and are very stable against aggregation because of the surfactant molecules attached to the surface. The method has been improved to obtain MNPs with $<d> > 12$ nm by growing previously synthesized MNPs as seeds (~ 10 nm) and repeating the synthesis protocol to further increase the final particle size. In this way, the synthesis of particles up to <d> = 25 nm has been reported.

### Oxidative hydrolysis method

This method, first reported by Matijevic et al. **[16]**, is based on the precipitation of an iron salt ($FeSO_4$) in basic media (NaOH) in the presence of a mild oxidant. It was later improved for specific applications **[17, 18]**. Two different approaches have been reported to coat MNPs, including in-situ and after–synthesis with organic polymers such as poly L-lysine (PLL), polyethylene glycol (PEG) and PEI. In the first approach, the MNPs are coated during the synthesis, while the post-synthesis coating method consists on grafting the polymer or surfactant onto the magnetic particles once synthesized.

## MAGNETOFECTION IN CELLS

As already stated, magnetofection was initially developed mainly to enhance gene transfer in cell culture, a simpler and more easily controllable scenario than in vivo models. Magnetofection in cell lines not only facilitated the optimization of protocols and MNP formulations but it also provided evidence for some unique capabilities of this approach. Progressively, an increasing number of publications combining magnetofection in cell culture and in experimental animals are beginning to emerge. This section will review studies exclusively dealing with magnetofection in cells, leaving for the next section the consideration of reports documenting in vivo studies.

### Neuronal and glial cells

Since neurons are sensitive to cytotoxicity and generally difficult to transfect by conventional methods, there is a growing interest in developing MNP formulations and magnetofection protocols suitable for neuronal cell cultures. One such protocol was optimized for transfection of cDNA and RNA interference (short hairpin RNA (shRNA)) into rat hippocampal neurons (embryonic day 18/19) cultured for several hours to 21 days. The protocol allowed double-transfection of DNA into a small subpopulation of hippocampal neurons (GABAergic interneurons), and achieved long-lasting expression of DNA and shRNA constructs without interfering with neuronal differentiation **[19]**. A specific MNP formulation called NeuroMag, which uses particles ranging in size from 140 to 200 nm and possessing a positive zeta potential, has recently been reported to significantly enhance reporter gene transfer in mouse neural stem cell (NSC) cultures without showing significant levels of toxicity **[20].** Magnetofection has been also used for effective gene transfer in cultures of multipotent rat neural precursor cells and rat oligodendrocyte precursors **[21, 22].** In primary cultures of rat hypothalamic neurons, magnetofection was used to transfect the CG and CA alleles of an enhancer sequence related to galanin expression **[23].** Magnetofection in mouse embryonic motor neurons was used to transfect a plasmid encoding the gene for a fluorescent protein fused to the spinal muscular atrophy-disease protein Smn. With this approach it was demonstrated that Smn is actively transported along axons of live primary motor neurons. Furthermore, magnetofection was also implemented for gene knockdown using shRNA-bound constructs **[24]**.

**Endothelial and epithelial cells**

Magnetofection has been reported to potentiate gene delivery to cultured primary endothelal cells and to human umbilical vein endothelial cells (HUVEC). Thus, up to a 360-fold increase in luciferase gene transfer was achieved by magnetofection as compared to various conventional transfection methods **[25]**. Biodegradable polylactide-based MNPs, surface-modified with the D1 domain of CAR as an affinity linker, have been affinity bound to a RAd expressing GFP and used to implement magnetofection in cultured endothelial and smooth muscle cells. This strategy yielded a stable MNP-RAd association that displayed efficient gene delivery and rapid cell binding kinetics in the presence of a magnetic field. Multiple regression analysis suggested that the mechanism by which the complex transduces the cells is different from that of naked adenoviruses **[26]**. More recently, the development of MNPs coated with PEG and with covalently linked branched PEI (bPEI), has been reported. In HUVEC cultures, nonviral vector-hybrid MNP complexes exhibited highly efficient magnetofection, even in serum conditioned media **[27]**.

In tissue engineering a major challenge comes from insufficient formation of blood vessels in implanted tissues. One approach to overcome this problem has been the production of angiogenic cell sheets using a combination of two techniques namely, magnetic cell accumulation and magnetofection with magnetite cationic liposomes (MCLs) coupled to a retroviral vector expressing vascular endothelial growth factor (VEGF). VEGF magnetofection in a monolayer of mouse myoblast C2C12 cells increased transduction efficiency by 6.7-fold compared with a conventional method. Then, MCL-labeled cells were accumulated in the presence of a magnetic field to promote the spontaneous formation of multilayered cell sheets. When these sheets were subcutaneoulsy grafted in nude mice they produced thick tissues displaying a high-cell density **[28]**. Magnetic field- and ultrasound-aided delivery of the gene for VEGF(165) to oversized ischemic rat skin flaps was implemented using magnetic lipospheres (magnetobubbles) loaded with the corresponding cDNA. This approach increased the survival and perfusion of flaps grafted in rats **[29]**.

Topical application of DNA vector complexes to the airways faces specific extracellular barriers. In particular, short contact time of complexes with the cell surface caused by mucociliary clearance hinders cellular uptake of complexes. In order to overcome this limitation magnetofection of the luciferase gene was assessed in permanent (16HBE14o-) and primary airway epithelial cells (porcine and human) as well as in native porcine airway epithelium ex vivo. Transfection efficiency and dose-response relationship of the luciferase gene revealed that magnetofection enhanced transfection efficiency in both, permanent and primary airway epithelial cells. Magnetofection also induced significant transgene expression at very short incubation times in the ex vivo airway epithelium organ model **[30]**. Magnetically guided lentiviral-mediated transduction of bronchial epithelial cells was also reported to induce efficient reporter (GFP) gene delivery **[31]**. In another study, MNPs complexed to Lipofectamine 2000 or cationic lipid 67/plasmid DNA (pDNA) liposome complexes were reported to be highly effective for gene delivery in airway epithelial cell cultures but less effective than pDNA alone when applied in the murine nasal epithelium in vivo. The latter result is likely to be a consequence of the significant precipitation of the complexes observed in vivo **[32].**

**Tumor and embryonic cells**

Hexanoyl chloride-modified chitosan (Nac-6) stabilized iron oxide nanoparticles (Nac-6-IOPs) were used in the CAR(!) human leukemia K562 cell line for viral gene (RAd-LacZ) delivery via magnetofection. For this complex the authors reported effective magnetofection results in vitro and in vivo **[33].** In a recent study, the transfection efficiency (percentage of transfected cells) and therapeutic potential (potency of insulin-like growth factor–1 receptor (IGF-1R) knockdown) of liposomal magnetofection of plasmids expressing GFP and shRNAs targeting IGF-1R (pGFPshIGF-1Rs) was assessed in A549 lung adenocarcinoma cells and in tumor-bearing mice. This method was reported to achieve a 3-fold improvement in GFP expression as compared to lipofection using Lipofectamine 2000. In vitro, IGF-1R specific-shRNA transfected by lipofection and by magnetofection inhibited IGF-1R protein by 56.1±6% and 85.1±3%, respectively. In vivo delivery efficiency of the pGFPshIGF-1R plasmid into the tumor was significantly higher in the liposomal magnetofection group than in the lipofection group **[34]**.

Magnetofection of cDNA constructs and shRNA into mouse genital ridge tissue was implemented as a means of gain-of-function and loss-of-function analysis, respectively. Ectopic expression of Sry induced female-to-male sex-reversal, whereas knockdown of Sox9 expression caused male-to-female sex-reversal, consistent with the known functions of these genes. Also, ectopic expression of Tmem184a, a gene of unknown function, in female genital ridges, resulted in failure of gonocytes to enter meiosis. These results suggest that magnetofection may constitute a suitable tool for the study of gene function in a broad range of developing organs and tissues **[35]**.

## THERAPEUTIC POTENTIAL OF MAGNETIC GENE TARGETING IN VIVO

### Cancer

Cancer has been a major target disease for gene therapy since the early days of this technology. Currently, a high number of experimental and clinical studies are under way using a wide variety of approaches to deliver the cDNA of choice to the tumor cells. The nature of these approaches depends on the biological characteristics of the tumor to be treated and include the delivery of genes for immunomodulatory molecules, suicide genes, tumor suppressor genes, oncolytic genes and antiangiogenic genes, among others **[for a review see 36; also see 37, 38]**. In many instances the above approaches involve invasive procedures when local administration of the vector is required. If the therapeutic gene vector is administered intravenously (IV), high doses need to be injected to compensate for dilution of the vector in the circulation. This also leads to spreading of the vector throughout the body, with lungs, liver and kidneys accumulating substantial levels of the vector. These limitations are also faced by pharmacological approaches using anticancer drugs, which usually are significantly toxic for healthy organs. This prompted the development of magnetic carriers and MDT whose two main goals are, a) to reduce the invasiveness of drug administration and b) to generate a "magnetic cage" in the target area so that the magnetic carriers are trapped and concentrated there. In this way lower doses of the antitumor drug would be necessary for achieving therapeutically effective intratumor levels. Magnetic trapping would also minimize drug dissemination to the rest of the organism.

Magnetic carriers were first used to target cytotoxic drugs (doxorubicin) to sarcoma tumors implanted in rat tails **[39].** The initial results were encouraging, showing a total remission of the sarcomas compared to no remission in another group of rats which were administered with ten times the dose but without magnetic targeting. Since that study, success in cytotoxic drug delivery and tumor remission has been reported by several groups using animal models including swine **[40, 41],** rabbits **[42]** and rats **[43, 44, 45].** This technique has also been employed to target cytotoxic drugs to brain tumors which are particularly difficult targets due to the fact that the drug must cross the blood brain barrier (BBB). It was reported that microparticles 1–2 $\mu$m in diameter could be concentrated in an intracerebral rat glioma **[44].** Although the concentration of the particles in the tumor was low it was significantly higher than in controls injected with nonmagnetic particles. Better results were achieved in these tumors employing 10–20 nm MNP particles **[45]**. Electron microscopic analysis revealed the presence of MNPs in the interstitial space of the tumors but not in normal brain tissue where MNPs were only found in the vasculature. In another study, MDT in rat brain tumors achieved some degree of success only when the BBB was disrupted immediately prior to particle injection **[46].**

There have been a few trials of MDT in humans although none of them has been followed up and currently no major pharmaceutical company has undertaken the development of magnetic drug formulations. A Phase I clinical trial demonstrated that the infusion of ferrofluids was well tolerated in most of the 14 patients studied **[2]**. In addition, the authors reported that the magnetic particles were successfully directed in advanced sarcomas without associated organ toxicity. Multi-center Phase I and II, MDT clinical trials for hepatocellular carcinomas, employing magnetic microspheres to which doxorubicin hydrochloride had been adsorbed, revealed promising preliminary results **[47]**. Although clinical application of MDT still faces technical limitations, pre-clinical and experimental results indicate that it is possible to overcome some of the reported problems by means of technical improvements of the magnetic delivery systems **[2, 48].** Clearly**,** the prospect of using magnetic carrier-gene vector complexes emerges as a promising avenue for cancer gene therapy. This approach has been used to implement immunostimulating gene therapy in domestic cats with clinical diagnosis of fibrosarcoma. Different doses of a plasmid harboring the gene for either feline Interferon-γ, feline interleukin-2 or feline granulocyte-macrophage colony stimulating factor (felGM-CSF) were complexed with MNPs. The complexes were intratumorally injected and an external magnetic field was applied. The treatment was well-tolerated by most of the animals **[49]**. In a follow up phase I trial, preoperative felGM-CSF gene therapy had favorable results as assessed by the rate of recurrence in treated versus (surgery-only) control cats **[50].** More recently, Tresilwised *et al.* **[7]** examined the potential of boosting the efficacy of the oncolytic adenovirus dl520 by associating it with MNP and performing magnetic field-guided infection in multidrug-resistant cancer cell cultures and in a murine xenograft model. Upon intratumoral injection and application of a gradient magnetic field, magnetic virus complexes exhibited a stronger oncolytic effect than adenovirus alone**.**

**Neurological diseases**

Gene transfer to the central nervous system (CNS) poses significant challenges due to both the relative inaccessibility of the brain and the extraordinary complexity of CNS structures. On the other hand, this approach offers unique advantages for the long-term delivery of neurotrophic factors to specific CNS regions affected by neurodegenerative processes and other neurological pathologies. Although the documented results for gene therapy in animal models of Parkinson's Disease **[51-54],** Alzheimer Disease **[55, 56]** and other neurological pathologies **[57, 58]** are promising, up to now the only way to administer the therapeutic vectors is via stereotaxic injections in the target brain areas. The invasiveness of this procedure significantly limits its eventual implementation in human patients.

The technology for magnetic field-assisted gene delivery has now advanced to a point from where it seems feasible to implement minimally invasive gene therapy strategies for the brain. This approach, which combines MDT and magnetofection, appears particularly suitable for pathologies in which the affected brain regions can be reached by the therapeutic molecules when they are released into the cerebrospinal fluid (CSF). In rats, it has been shown that adenoviral vectors injected intracerebroventricularly (ICV) efficiently transduce the ependymal cell layer and if they harbor the gene for a secreted peptide, it is released into the CSF **[59].** The ependymal route has been successfully used to implement cytokine-gene therapy in the CNS. In this case, ICV injection of a RAd vector expressing human interleukin IL-10 ameliorated disease signs in mice with active experimental autoimmune encephalomyelitis (EAE) **[60].** Furthermore, it is well-established that the delivery of genes encoding IL-10, IL-4, TGF-β, IFN-β, p55TNFR-Ig and p75TNFR-Ig into the CNS, is superior to IV administration of the same anti-inflammatory cytokines in the treatment of murine EAE **[61-63].** In aging rats, ICV implementation of IGF-I gene therapy ameliorated their deficient motor performance **[64].** Although the specific mechanisms that favor adenoviral transduction of ependymal cells are unknown, this route of gene delivery has numerous advantages including the ability to increase the levels of a transgenic therapeutic protein throughout many regions of the CNS. It also avoids possible side effects of pharmacologically high circulating levels of therapeutic molecules after peripheral administration.

The ependymal route has been recently used to implement MGT in rodent embryos. Thus, a RAd vector tagged with MNPs was ICV injected in mouse embryos *in vitro* and *in vivo*. By applying an external magnetic field to a limited area of the head of the embryos, transgene delivery was restricted to that region **[65]**. The same route could be exploited to implement minimally invasive therapeutic gene delivery in the adult rodent brain by ICV administration of MNP-viral vector complexes at distal sites and subsequent magnetic trapping of the complexes at the target brain region by means of a properly focused external magnetic field. There are a number of suitable adult animal models available for trial **[66],** one of them being the aging female rat. In effect, it is well-established that in the female rat, the hypothalamic dopaminergic (DA) neurons which exert a tonic inhibitory control on prolactin secretion, become dysfunctional with age **[67].** A significant reversal of chronic hyperprolactinemia and hypothalamic DA neuron dysfunction was achieved by neurotrophic factor gene therapy in the hypothalamus of aged female rats **[68, 69].** In these studies the therapeutic viral vectors were injected into the hypothalamic parenchyma. It is proposed that similar results could be achieved by a less invasive approach involving the injection of MNP-therapeutic viral vector complexes in the cisterna magna and subsequently concentrating them by magnetic trapping in the third ventricle **(Fig. 5).** To reach the third ventricle from the cisterna magna the magnetic complexes need to travel counterflow. Since CSF flow velocities are over ten times lower than arterial blood flow velocities (0.4 cm/s **[70]** versus 5 cm/s **[71]**, respectively), the strength of the magnetic field to be applied in order to ovecome CSF counterflow force remains within the capacity of cylindrically or conically shaped Nd-Fe-B permanent magnets. If successful, this proof-of-concept approach could be extended to other regions of the brain.

**Myocardial Infarction**

Heart failure remains as one of the major causes of morbidity and mortality throughout the world, worsening as the population ages. The development of the coronary bypass implant technique and its implementation in human patients **[72]** represented a major achievement for the surgical treatment of myocardial ischemia. The search for less invasive approaches led to the development of the nonsurgical technique known as percutaneous transluminal coronary angioplasty (PTCA) **[73]** which has revolutionized the treatment of acute coronary failure, preventing or significantly reducing the consequences of myocardial infarction (MI). The subsequent development of drug-eluting stents has contributed to reduce the incidence of post-angioplasty restenosis due to proliferation and migration of medial and intimal smooth muscle cells (SMC) in the treated artery, a significant problem with early bare metal stents. Coronary stenting technology has made it conceivable the clinical implementation of cardiovascular gene therapy (for a general review on cardiovascular gene and cell therapy see **[74]**). Viral vectors harboring genes for angiogenic, myotrophic or anti-proliferative factors can now be delivered in animal models by the use of viral vector-eluting stents. Such strategies have been reported in rabbit vascular injury models **[75, 76]**. One of the current targets of experimental gene therapy approaches is to prevent restenosis by local delivery of genes encoding SMC antiproliferative factors. For instance, RAd-mediated overexpression of the cyclin/cyclin dependent kinase (CDK) inhibitor

p21, was used to inhibit neointima formation in a rat model of balloon angioplasty **[77]**. In another gene therapy approach, transcription decoys using a consensus-binding sequence for transcription factor E2F inhibited smooth muscle proliferation in a model of rat carotid injury **[78].**

Another aim of cardiovascular gene therapy is to stimulate myocardial angiogenesis in the post-MI heart. In a swine model of pacing-induced congestive heart failure, intramyocardial injection of RAd-$VEGF_{121}$ increased myocardial perfusion and enhanced its function **[79].** At clinical level, in a phase II randomized controlled trial using RAd-$VEGF_{121}$, there was improvement in exercise-induced ischemia in patients that received intramyocardial delivery of the therapeutic vector **[80]**. In patients with previous MI or angina, RAd-mediated delivery of $VEGF_{165}$ or fibroblast growth factor (FGF)-4 was reported to be effective in increasing myocardial perfusion **[81, 82].** Another important angiogenic candidate factor for myocardial gene therapy is IGF-I. Thus, in a rat model of MI, local IGF-I gene delivery by an adeno-associated viral vector (AAV) rendered sustained transduction and improved cardiac function post-MI **[83]**.

The combination of intra-arterial gene vector delivery by coronary catheterization with MGT could further improve the effectiveness of post-MI gene therapy. MDT studies in mice demonstrated that an external stationary magnetic field (**ΛB=** 200mT/cm) focused on the lung could achieve a significant magnetic field and field gradient in the heart (112 mT and 90 mT/cm, respectively) increasing the bioavailability of doxorubicin-magnetite nanoparticle conjugates in the mouse lung **[84].** This suggests that in rodents, IV injection of MNP-gene vector complexes in the presence of a strong external magnetic field focused on the heart could achieve a significant concentration of the vector in the myocardium. In the first study to demonstrate *in vitro* and *in vivo* magnetically targeted gene delivery, magnetic microspheres were coated with an AAV2 encoding GFP or human α-1 anti-tripsin (AAT), using a cleavable heparin sulphate linker. The complexes induced increased gene delivery in C2 muscle cells and could be targeted by an external magnetic field. Increased gene delivery was achieved *in vivo* following intramuscular or IV injection of the complexes in mice **[85, 86]**. In these studies, IV injection of the complexes induced higher gene delivery to the heart (and other organs) than injection of the vector alone.

In spite of the promise these experimental studies offer, it is important to mention that in human patients, an external magnet placed over the chest would need to generate a very strong magnetic field in order to achieve in the heart, field gradients high enough as to prevent the arterial blood flow from washing away the MNP-vector complexes.

An alternative strategy to improve magnetic force is to insert a magnetizable coronary stent at the target site. Under the influence of an external magnetic field, the stent will create locally a high-gradient magnetic field. This procedure is termed implant-assisted magnetic drug targeting **[87-89].** The feasibility of this approach was suggested by a study in an isolated swine heart ventricle perfusion model carrying an intra-arterial stent coil fabricated from ferromagnetic stainless steel 430 wire and used to capture 100-nm diameter magnetite particles that mimicked magnetic drug carrier particles **[90]**. Implant-assisted targeting of magnetic particles under the influence of an external magnetic field has previously been verified through mathematical modeling **[91, 92]**, *in vitro* studies **[93]**, and *in vivo* studies in rat carotid arteries **[94, 95]** as a feasible method for localized drug delivery. An initial *in vivo* biocompatibility test in pigs, carried out by intravascular injection of the nanoparticles in a stented brachial artery, showed no short-term adverse effects. *In vitro* evaluation in a flow-through model proved that the magnetic nanoparticles were captured efficiently to the surface of a ferromagnetic coiled wire at the fluid velocities typical for human arteries. A preliminary test of tissue plasminogen activator (t-PA)-nanoparticle conjugates in a pig model suggested that the conjugates may be used for treatment of in-stent thrombosis in coronary arteries **[96]**.

The above studies are encouraging and suggest that MGT to the cardiovascular system could be a rewarding research avenue and that it merits to be explored further.

## CONCLUDING REMARKS

During the past two decades the biomedical applications of magnetic fields and MNPs have expanded remarkably due to the possibilities they open for noninvasive diagnostic and therapeutic approaches. In this context, the discovery that MNP-gene vector complexes can, in the presence of a magnetic field, greatly enhance gene transfer into cells and eventually allow the development of minimally invasive gene delivery approaches in vivo, is raising much interest in this emerging technology. Many of the studies reviewed here constitute important landmarks in the path towards a mature MGT technology.

In his seminal book, Engines of Creation **[97]**, KE Drexler defined nanotechnology as a manufacturing methodology based on the manipulation of individual atoms and molecules in order to construct complex structures, specified at the atomic level. In practice, Drexlerian nanotechnology remains as an embryonic discipline, with its practical implementation lying in the future. What is generally known as nanotechnology should be called applied nanoscience which is a discipline in active development. Drexlerian theorists imagine a medical branch of nanotechnology called nanomedicine **[98]**. This medical specialty will be based on the use of intelligent nanoinstruments or nanobots which after being injected into the bloodstream will survey the body searching for faulty cells, repairing them or destroying those beyond repair **(Fig. 6, left)**. These nanobots will be wirelessly controlled by external computers. **Figure 6 right,** diagrammatically represents a current MNP-adenovector complex. It could be considered as a gene delivery nanoinstrument. Its central component, the viral vector, has the capability to recognize and enter its target cells and deliver to them its therapeutic gene(s). To a certain extent it can also be wirelessly controlled, not by a computer, but by a magnetic field. Therefore, if Drexlerian nanomedicine becomes a reality in the future, perhaps these magnetic complexes will be considered as predecessors of therapeutic nanobots.

**ACKNOWLEDGEMENTS**

The authors are grateful to Ms. Yolanda Sosa for editorial and technical assistance and to Dr. Alicia Mattiazzi and Fracisco Sánchez for critical reading of the manuscript. Part of the work from our laboratory reviewed here was supported by NIH grant # R01AG029798-3, ANPCYT grant #PICT08-639 and CONICET grant PIP2378 to RGG and by the EU-Argentinean EULANEST grant NEURONANO-31 to GFG and RGG. PCR, CBH and RGG are Argentine Research Council (CONICET) career researchers. JIS is a CONICET doctoral fellow. The authors have no conflict of interests.

**REFERENCES**

[1] Widder KJ, Senyel AE, Scarpelli GD. Magnetic microspheres: A model system of site specific drug delivery in vivo.Proc Soc Exp Biol Med 1978;158: 141-46.

[2] Lübbe AS, Bergemann C, Riess H, *et al.*Clinical experiences with magnetic drug targeting: a phase I study with 4'-epidoxorubicin in 14 patients with advanced solid tumors. Cancer Res 1996; 56 : 4686-93.

[3] Lübbe AS, Alexiou C, Bergemann C. Clinical applications of magnetic drug targeting. J Surg Res 2001; 95 : 200-6.

[4] Scherer F, Anton M, Schillinger U, *et al.* Magnetofection: enhancing and targeting gene delivery by magnetic force in vitro and in vivo. Gene Ther 2002; 9: 102-9.

[5] Plank C, Zelphati O, Mykhaylyk O. Magnetically enhanced nucleic acid delivery. Ten years of magnetofection - Progress and prospects. Adv Drug Deliv Rev (2011, in press) **doi:10.1016/j.addr.2011.08.002**

[6] Huth S, Lausier J, Gersting SW, *et al.*Insights into the mechanism of magnetofection using PEI-based magnetofectins for gene transfer. J Gene Med 2004; **6:** 923-36.

[7] Tresilwised N, Pithayanukul P, Mykhaylyk O, *et al.* Boosting oncolytic adenovirus potency with magnetic nanoparticles and magnetic force. Mol Pharm 2010;7 :1069-89.

[8] Kamei K, Mukai Y, Kojima H, *et al.* Direct cell entry of gold/iron-oxide magnetic anoparticles in adenovirus mediated gene delivery. Biomaterials 2009; 30 **:** 1809-14.

[9] Scherer F, Plank C. Magnetofection: Using magnetic particles and magnetic force to enhance and to target nucleic acid delivery. In: Smyth Templeton N, Ed. Gene and Cell Therapy: Therapeutic mechanisms and strategies; 3rd Ed. Boca Raton, FL: CRC Press 2009; pp.379-404.

[10] Mykhaylyk O, Antequera YS, Vlaskou D, Plank C. Generation of magnetic nonviral gene transfer agents and magnetofection in vitro. Nat Protoc 2007; 2: 2391-411.

[11] McBain SC, Yiu HHP, Dobson J. Magnetic nanoparticles for gene and drug delivery .Int J Nanomedicine 2008; 3(2): 169–180.

[12] Sun, SH, Zeng H, Robinson DB, *et al.* Monodisperse MFe2O4 (M = Fe, Co, Mn) nanoparticles. J Am Chem Soc 2004; 126 : 273-79.

[13] |Robinson DB, Persson HH, Zeng H *et al*. DNA-functionalized MFe2O4 (M = Fe, Co, or Mn) nanoparticles and their hybridization to DNA-functionalized surfaces. Langmuir 2005; 21: 3096-103.

[14] Goya, GF, Lima Jr E, Arelaro AD, *et al*. Magnetic Hyperthermia With Fe3O4 Nanoparticles: The Influence of Particle Size on Energy Absorption. IEEE Trans Magnetics 2008; 44: 4444-47.

[15] 25. Vargas JM, Zysler RD. Tailoring the size in colloidal iron oxide magnetic nanoparticles. Nanotechnology 2005; 16: 1474-76.

[16] Matijevic E, Scheiner P. Ferric hydrous oxide sols .3. preparation of uniform particles by hydrolysis of fe(iii)-chloride, fe(iii)-nitrate, and fe(iii)-perchlorate solutions. J Colloid Interface Sci 1978; 63: 509-24.

[17] Verges MA., Costo R, Roca AG, *et al*. Uniform and water stable magnetite nanoparticles with diameters around the monodomain-multidomain limit. J Phys D-Appl Phys 2008; 41: 134003

[18] Gonzalez-Fernandez MA., Torres T, Verges R, et al. Magnetic nanoparticles for power absorption: Optimizing size, shape and magnetic properties. J Solid St Chem 2009; 182: 2779-84.

[19] Buerli T, Pellegrino C, Baer K, *et al*. Efficient transfection of DNA or shRNA vectors into neurons using magnetofection; Nat Protocols 2007; 2: 3090-101.

[20] Sapet C, Laurent N, de Chevigny A, *et al.* High transfection efficiency of neural stem cells with magnetofection. Biotechniques 2011; **50:** 187-9.

[21] Pickard MR, Barraud P, Chari DM. The transfection of multipotent neural precursor/stem cell transplant populations with magnetic nanoparticles. Biomaterials 2011; 32**:** 2274-84.

[22] Jenkins SI, Pickard MR, Granger N, Chari DM. Magnetic Nanoparticle-Mediated Gene Transfer to Oligodendrocyte Precursor Cell Transplant Populations Is Enhanced by Magnetofection Strategies. ACS Nano 2011 Jul 13. [Epub ahead of print]

[23] Davidson S, Lear M, Shanley L, *et al.* Differential Activity by Polymorphic Variants of a Remote Enhancer that Supports Galanin Expression in the Hypothalamus and Amygdala: Implications for Obesity, Depression and Alcoholism; Neuropsychopharmacology 2011 Jun 29. doi: 10.1038/npp.2011.93. [Epub ahead of print]

[24] Fallini C, Bassell GJ, Rossoll W. High-efficiency transfection of cultured primary motor neurons to study protein localization, trafficking, and function. Mol Neurodegener 2010 5**:** 17.

[25] Krötz F, Sohn HY, Gloe T, Plank C, Pohl U. Magnetofection potentiates gene delivery to cultured endothelial cells J Vasc Res 2003; 40 : 425-34.

[26] Chorny M, Fishbein I, Alferiev I, Levy RJ. Magnetically responsive biodegradable nanoparticles enhance adenoviral gene transfer in cultured smooth muscle and endothelial cells. Mol Pharm 2009; 6: 1380-87.

[27] Namgung R, Singha K, Yu MK, *et al.* Hybrid superparamagnetic iron oxide nanoparticle-branched polyethylenimine magnetoplexes for gene transfection of vascular endothelial cells. Biomaterials 2010 ; 31: 4204-13.

[28] Akiyama H, Ito A, Kawabe Y, Kamihira M. Genetically engineered angiogenic cell sheets using magnetic force-based gene delivery and tissue fabrication techniques. Biomaterials 2010; 31: 1251-59.

[29] Holzbach T, Vlaskou D, Neshkova I, *et al.* Non-viral VEGF(165) gene therapy--magnetofection of acoustically active magnetic lipospheres ('magnetobubbles') increases tissue survival in an oversized skin flap model. J Cell Mol Med 2010; 14 : 587-99.

[30] Gersting SW, Schillinger U, Lausier J, *et al.* Gene delivery to respiratory epithelial cells by magnetofection.; J Gene Med 2004; 6 : 913-922.

[31] Orlando C, Castellani S, Mykhaylyk O, *et al.* Magnetically guided lentiviral-mediated transduction of airway epithelial cells. J Gene Med 2010; 12 :747-54.

[32] Xenariou S, Griesenbach U, Ferrari S, *et al.* Using magnetic forces to enhance non-viral gene transfer to airway epithelium in vivo. Gene Ther 2006; 13: 1545-52.

[33] Bhattarai SR, Kim SY, Jang KY, et al. N-hexanoyl chitosan-stabilized magnetic nanoparticles: enhancement of adenoviral-mediated gene expression both in vitro and in vivo. Nanomedicine 2008 ; 4 **:** 146-54.

[34] Wang C, Ding C, Kong M, *et al.* Tumor-targeting magnetic lipoplex delivery of short hairpin RNA suppresses IGF-1R overexpression of lung adenocarcinoma A549 cells in vitro and in vivo. Biochem Biophys Res Commun 2011; 410 : 537-42.

[35] Svingen T, Wilhelm D, Combes AN, *et al.* Ex vivo magnetofection: a novel strategy for the study of gene function in mouse organogenesis. Dev Dyn 2009; 238 **:** 956-64.

[36] Brand, K. Gene therapy for cancer; In: Smyth Templeton N, Ed. Gene and Cell Therapy: Therapeutic mechanisms and strategies; 3rd Ed. Boca Raton FL, CRC Press 2009; pp. 761-99.

[37] Verreault M, Webb MS, Murray S,Ramsay EC, Bally MB. Gene silencing in the development of personalized cancer treatment: The targets, the agents and the delivery systems . Curr Gene Ther 2006; 6 ( 4) : 505-533.

[38] Sell, S. Potential gene therapy strategies for cancer stem cells. Curr Gene Ther 2006; 6 (5): 579-591.

[39] Widder KJ, Morris RM, Poore GA, Howard DP, Senyei AE. Selective targeting of magnetic albumin microspheres containing low-dose doxorubicin—total remission in Yoshida sarcoma-bearing rats. Eur J Cancer Clin Oncol 1983;19 : 135-39

[40] Goodwin S, Peterson C, Hob C, Bittner C . Targeting and retention of magnetic targeted carriers (MTCs) enhancing intra-arterial chemotherapy. J Magn Magn Mater 1999;194: 132–9

[41] Goodwin SC , Bittner CA, Peterson CL, Wong G .Single-dose toxicity study of hepatic intra-arterial infusion of doxorubicin coupled to a novel magnetically targeted drug carrier. Toxicol Sci.2001; 60 **:** 177–83

[42] Alexiou C, Arnold W, Klein RJ, *et al.* Locoregional cancer treatment with magnetic drug targeting. Cancer Res 2000; 60 : 6641–8

[43] Lübbe AS, Bergemann C, Brock J, McClure DG. Physiological aspects in magnetic drug-targeting . J. Magn Magn Mater 1999; 194: 149–55

[44] Pulfer SK, Gallo JM. Enhanced brain tumor selectivity of cationic magnetic polysaccharide microspheres J. Drug Targeting 1999; 6 **:** 215–28

[45] Pulfer SK, Ciccotto SL, Gallo JM. Distribution of small magnetic particles in brain tumor-bearing rats J Neuro-Oncol 1999; 41: 99–105.

[46] Mykhaylyk O, Cherchenko A, Ilkin A. Glial brain tumor targeting of magnetite nanoparticles in rats J Magn Magn Mater 2001; 225 : 241–7

[47] Wilson RW; A Phase I/II Trial of Hepatic Delivery of Doxorubicin Adsorbed to Magnetic Targeted Carriers in Patients with HCC; 28th Annual Scientific Meeting of the Society of Interventional Radiology, Salt Lake City, Utah, 2003 (Abstract); http://www.zeropresence.com/ferx/press03-31-03.htm

[48] Gallo JM, Häfeli U. Preclinical experiences with magnetic drug targeting: tolerance and efficacy and clinical experiences with magnetic drug targeting: a phase I study with 4'-epidoxorubicin in 14 patients with advanced solid tumors. Cancer Res 1997; 57**:** 3063–4.

[49] Jahnke A, Hirschberger J, Fischer C, et al. Intra-tumoral gene delivery of feIL-2, feIFN-gamma and feGM-CSF using magnetofection as a neoadjuvant treatment option for feline fibrosarcomas: a phase-I study. J Vet Med A Physiol Pathol Clin Med 2007; 54**:** 599-606.

[50] Hüttinger C, Hirschberger J, Jahnke A, *et al.* Neoadjuvant gene delivery of feline granulocyte-macrophage colony-stimulating factor using magnetofection for the treatment of feline fibrosarcomas: a phase I trial. J Gene Med 2008; 10**:** 655-67.

[51] Choi-Lundberg DL, Lin Q, Chang YN, *et al.* Dopaminergic neurons protected from degeneration by GDNF gene therapy. Science 1997; 275: 838-41.

[52] Connor B, Kozlowski DA, Schallert T, Tillerson JL , Davidson BL, Bohn MC Differential effects of glial cell line-derived neurotrophic factor (GDNF) in the striatum and substantia nigra of the aged Parkinsonian rat. Gene Ther 1999; 6: 1936-51.

[53] Kordower JH, Emborg ME, Bloch J, *et al.* Neurodegeneration prevented by lentiviral vector delivery of GDNF in primate models of Parkinson's disease. Science 2000; 290: 767-73.

[54] Carlsson T, Bjorklund T, Deniz K. Restoration of the striatal dopamine synthesis for Parkinson's disease: viral vector-mediated enzyme replacement strategy Curr Gene Ther 2007; 7 (2): 109-120.

[55] Tuszynski MH, Thal L, Pay M, *et al.* A phase 1 clinical trial of nerve growth factor gene therapy for Alzheimer disease. Nat Med 2005; 11: 551-5.

[56] Hong CS, Goins WF, Goss JR, Burton EA, Glorioso JC. Herpes simplex virus RNAi and neprilysin gene transfer vectors reduce accumulation of Alzheimer's disease-related amyloid-beta peptide in vivo. Gene Ther 2006 ; 13: 1068-79.

[57] Frisella WA, O'Connor LH, Vogler CA. Intracranial injection of recombinant adeno-associated virus improves cognitive function in a murine model of mucopolysaccharidosis type VII. Mol Ther 2001; 3: 351-8.

[58] Machida Y, Okada T, Kurosawa M, Oyama F, Ozawa K, Nukina N. rAAV-mediated shRNA ameliorated neuropathology in Huntington disease model mouse. Biochem Biophys Res Commun. 2006 ; 343 : 190-7

[59] Hereñu CB, Sonntag WE, Morel GR, Portiansky EL, Goya RG. The ependymal route for insulin-like growth factor-1 gene therapy in the brain. Neuroscience 2009; 163: 442-7.

[60] Cua DJ, Hutchins B, LaFace DM, Stohlman SA, Coffman RL .Central nervous system expression of IL-10 inhibits autoimmune encephalomyelitis. J Immunol 2001; 166: 602–8.

[61] Triantaphyllopoulos K, Croxford J, Baker D, Chernajovsky Y. Cloning and expression of murine IFN-β and a TNF antagonist for gene therapy of experimental allergic encephalomyelitis. Gene Ther 1998; **5:** 253-63.

[62] Croxford JL, Triantaphyllopoulos K, Podhajcer OL, Feldmann M, Baker D, Chernajovsky Y. Cytokine gene therapy in experimental allergic encephalomyelitis by injection of plasmid DNA–cationic liposome complex into the central nervous system. J Immunol 1998; 160: 5181–7.

[63] Croxford JL, Triantaphyllopoulos KA, Neve RM, Feldmann M, Chernajovsky Y, Baker D. Gene therapy for chronic relapsing experimental allergic encephalomyelitis using cells expressing a novel soluble p75 dimeric TNF receptor. J Immunol 2000; 164: 2776–81.

[64] Nishida F, Morel GR, Hereñú CB, Schwerdt JI, **Goya RG** and Portiansky EL. Restorative effect of intracerebroventricular Insulin-like Growth Factor-I gene therapy on motor performance in aging rats, Neuroscience 2011; 177: 185-206.

[65] Hashimoto M, Hisano Y. Directional gene-transfer into the brain by an adenoviral vector tagged with magnetic nanoparticles. J Neurosci Methods 2011; 194: 316-20 .

[66] Hereñú CB, Brown OA, Sosa YE, *et al***.** The neuroendocrine system as a model to evaluate experimental gene therapy. Curr Gene Ther 2006; 6**:** 125-9.

[67] Sánchez HL, Silva LB, Portiansky EL, Goya RG, Zuccolilli GO. Impact of very old age on hypothalamic dopaminergic neurons in the female rat: A morphometric study. J Comp Neurol 2003; 458: 319-25.

[68] Hereñú CB, Cristina C, Rimoldi OJ, *et al*. Restorative effect of Insulin-like Growth Factor-I gene therapy in the hypothalamus of senile rats with dopaminergic dysfunction. Gene Therapy 2007; 14 : 237-45.

[69] Morel GR, Sosa YE, Bellini MJ, *et al*. Glial cell line-derived neurotrophic factor gene therapy ameliorates chronic hyperprolactinemia in senile rats. Neuroscience 2010; 167: 946-53.

[70] Hazel RD, McCormack EJ, Miller J, *et al*.Measurement of Cerebrospinal Fluid Flow in the Aqueduct of a Rat Model of Hydrocephalus. Proc Intl Soc Mag Reson Med 2006; 14: 30 .

[71] Matsievskiy DD, Konorova IL, Lebedeva MA. Transcranial evaluation of blood flow velocity in the basilar artery in rats by high frequency ultrasonic Doppler technique. Bull Exp Biol Med 2009; 148: 568-71.

[72] Favaloro RG. Saphenous vein autograft replacement of severe segmental coronary artery occlusion: operative technique. Ann Thorac Surg 1968; 5: 334-39.

[73] Gruntzig A. Transluminal dilatation of coronary-artery stenosis. Lancet 1978; 1: 263.

**[74]** Staudacher DL, Flugelman MY. Cell and Gene Therapies in Cardiovascular Disease with Special Focus on the No Option Patient. Curr Gene Ther 2006; 6 (6): 609-623.

[**75]** Walter DH, Cejna M, Diaz-Sandoval L, *et al.* Local gene transfer of phVEGF-2 plasmid by gene-eluting stents: an alternative strategy for inhibition of restenosis. Circulation 2004; 110: 36-45.

[76] Sharif F, Hynes SO, McMahon J, *et al.* Gene-eluting stents: comparison of adenoviral and adeno- associated viral gene delivery to the blood vessel wall in vivo. Hum Gene Ther 2006; 17: 741-50.

[77] Chang MW, Barr E, Lu MM, Barton K, Leiden JM. Adenovirus-mediated over-expression of the cyclin/cyclin-dependent kinase inhibitor, p21 inhibits vascular smooth muscle cell proliferation and neointima formation in the rat carotid artery model of balloon angioplasty. J Clin Invest 1995; 96 **:** 2260-68.

[78] Morishita R, Gibbons GH, Horiuchi M, *et al.* A gene therapy strategy using a transcription factor decoy of the E2F binding site inhibits smooth muscle proliferation in vivo. Proc Natl Acad Sci USA. 1995; 92: 5855-59.

[79] Leotta E, Patejunas G, Murphy G, *et al.* Gene therapy with adenovirus-mediated myocardial transfer of vascular endothelial growth factor 121 improves cardiac performance in a pacing model of congestive heart failure. J Thorac Cardiovasc Surg. 2002; 123 : 1101-13.

[80] Stewart DJ, Hilton JD, Arnold JM, *et al.* Angiogenic gene therapy in patients with nonrevascularizable ischemic heart disease: a phase 2 randomized, controlled trial of AdVEGF(121) (AdVEGF121) versus maximum medical treatment. Gene Ther 2006; **13 :** 1503-11.

[81] Hedman M, Hartikainen J, Syvänne M, *et al.* Safety and feasibility of catheter-based local intracoronary vascular endothelial growth factor gene transfer in the prevention of postangioplasty and in-stent restenosis and in the treatment of chronic myocardial ischemia: phase II results of the Kuopio Angiogenesis Trial (KAT). Circulation 2003; 107 **:** 2677-83.

[82] Grines CL, Watkins MW, Mahmarian JJ, *et al.* Angiogene GENe Therapy (AGENT-2) Study Group. A randomized, double-blind, placebo-controlled trial of Ad5FGF-4 gene therapy and its effect on myocardial perfusion in patients with stable angina. J Am Coll Cardiol 2003; 42: 1339-47.

[83] Dobrucki LW, Tsutsumi Y, Kalinowski L, *et al.* Analysis of angiogenesis induced by local IGF-1 expression after myocardial infarction using microSPECT-CT imaging.. J Mol Cell Cardiol 2010; 48:1071-79.

[84] Mykhaylyk OM, Dudchenko NO, Dudchenko AK. Pharmacokinetics of the doxorubicin magnetic nanoconjugate in mice. Effects of the nonuniform stationary magnetic field. Ukr Biokhim Zh 2005; 77: 80-92.

[85] Mah C, Zolotukhin I, Fraites TJ, Dobson J, Batich C, Byrne BJ . Microsphere-mediated delivery of recombinant AAV vectors *in vitro* and *in vivo.* Mol Ther 2000; 1 **:** S239

[86] Mah C, Fraites T J, Zolotukhin I, *et al.* Improved method of recombinant AAV2 delivery for systemic targeted gene therapy. Mol Ther 2002*;* 6 **:** 106–12

[87] Yellen BB, Forbes ZG, Halverson DS, *et al.* Targeted drug delivery to magnetic implants for therapeutic applications. J Magn Magn Mater 2005; 293: 647-54.

[88] Rosengart AJ, Kaminski MD, ChenH, Caviness PL, Ebner AD, Ritter JA. Magnetizable implants and functionalized magnetic carriers: a novel approach for noninvasive yet targeted drug delivery. J Magn Magn Mater 2005; 293: 633-38.

[89] Fernández- Pacheco R,Valdivia JG, Ibarra MR. Magnetic nanoparticles for local drug delivery using magnetic implants.Meth Mol Biol 2009; 544: 559-69.

[90] Avilés MO, Mangual JO, Ebner AD, Ritter JA. Isolated swine heart ventricle perfusion model for implant assisted-magnetic drug targeting; Int J Pharm 2008; 361: 202-8.

[91] Chen H, Ebner AD, Rosengart AJ, Kaminski MD, Ritter JA. Analysis of magnetic drug carrier particle capture by a magnetizable intravascular stent: 1. Parametric study with single wire correlation. J Magn Magn Mater 2004; 284: 181-94.

[92] Chen A, Ebner AD, Kaminski MD, Rosengart AJ, Ritter JA. Analysis of magnetic drug carrier particle capture by a magnetizable intravascular stent-2: parametric study with multi-wire two-dimensional model. J Magn Magn Mater 2005; 293: 616-32.

[93] **A**vilés MO, Chen H, Ebner AD, Rosengart AJ, Kaminski MD, Ritter JA. In vitro study of ferromagnetic stents for implant assisted-magnetic drug targeting. J Magn Magn Mater 2007; 311 : 306-11.

[94] Polyak B, Fishbein I, Chorny M, *et al.* High field gradient targeting of magnetic nanoparticle-loaded endothelial cells to the surfaces of steel stents. Proc Natl Acad Sci USA 2008; 105: 698-703.

[95] Chorny M, Fishbein I, Yellen BB, *et al*. Targeting stents with local delivery of paclitaxel-loaded magnetic nanoparticles using uniform fields. Proc Natl Acad Sci USA 2010 ; 107: 8346-51.

[96] Kempe H, Kempe M; The use of magnetite nanoparticles for implant-assisted magnetic drug targeting in thrombolytic therapy; Biomaterials 2010; 31: 9499-510.

[97] Drexler KE, Engines of creation.The coming era of nanotechnology. New York: Anchor Books 1986.

[98] Freitas RA, Jr., Nanomedicine, Vol. IIA; Biocompatibility. Austin TX: Landes Bioscience 2003.

**Table 1.** Values of saturation magnetization MS for different magnetic materials used as carriers in MDT and magnetofection.

| Material | $M_S$ (emu/g) † |
|---|---|
| Magnetite $Fe_3O_4$ | 90-92 |
| Maghemite $\gamma$-$Fe_2O_3$ | 84-88 |
| $CoFe_2O_4$ | ~75 |
| Iron ($\alpha$-Fe) | 217.9 |
| Cobalt | 162.7 |
| Nickel | 57.5 |

(† *values at room temperature*.)

## FIGURE LEGENDS

**Figure 1.- Diagrammatic representation of the magnetofection principle in cells.** MNPs are complexed to RAds and the complex is attracted to cells by a magnetic field. (Kindly provided by OZ Biosciences, Marseille, France, www.ozbiosciences.com).

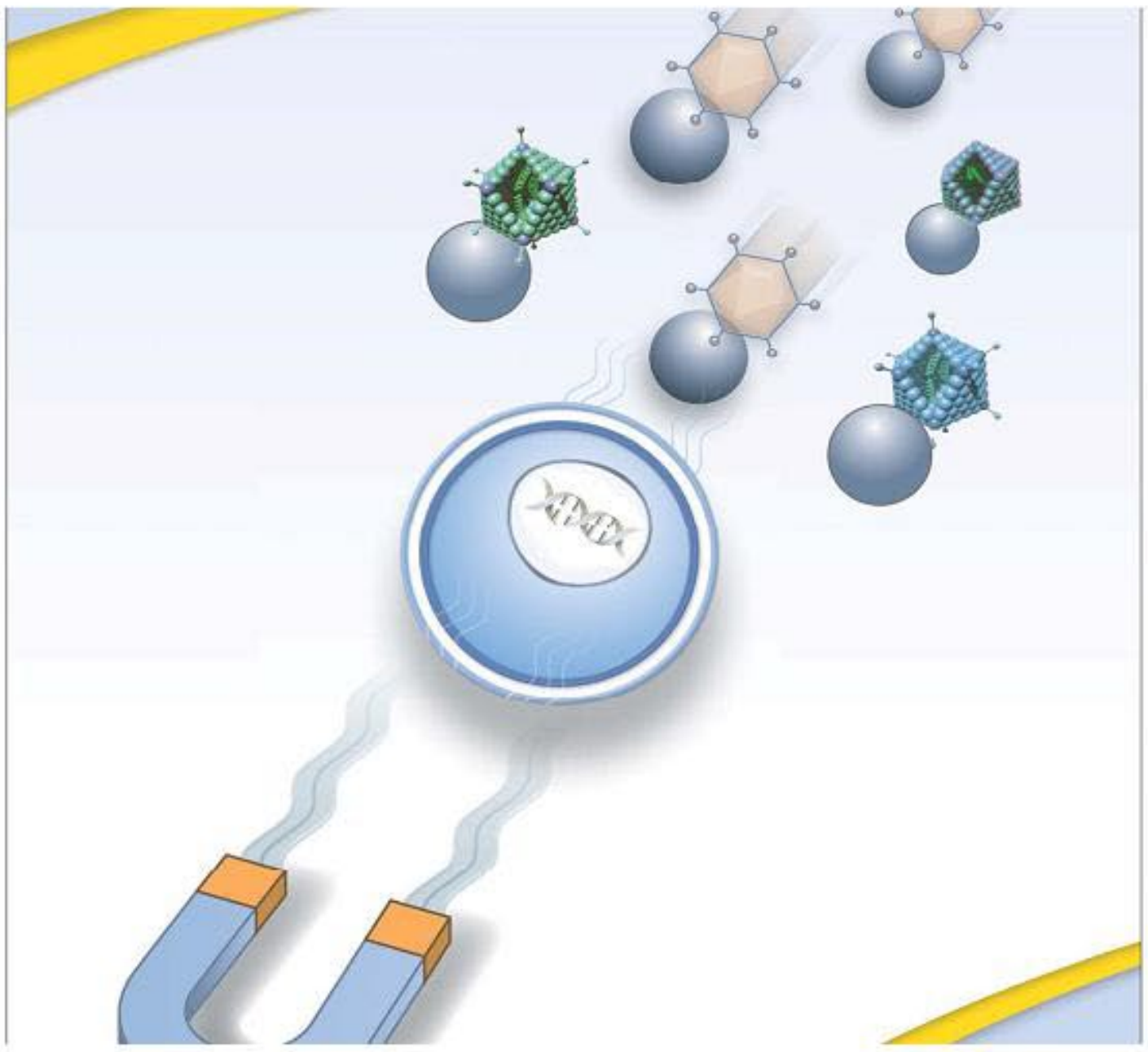

**Figure 2.- Magnetofection in 293 cells.-** Cell cultures were incubated with either $10^5$ pfu/well RAd-GFP alone (left image) or with 1 or 4 µl AdenoMag™ MNPs complexed to $10^5$ pfu/well RAd-GFP (center and right images, respectively). All cultures were exposed for 25 min to a magnetic field and images were taken 4 days afterwards. A higher number of transduced cells is evident in the cells incubated with the vector complexed to MNPs. The diagrams below are only intended to qualitatively illustrate the reader on the nature of the RAd-MNP complexes. They do not represent actual MNP/RAd-GFP ratios or complex structure. RAd-GFP, an adenoviral vector expressing the gene for green fluorescent protein. Obj. 20X **(Goya, RG, et al., unpublished data).**

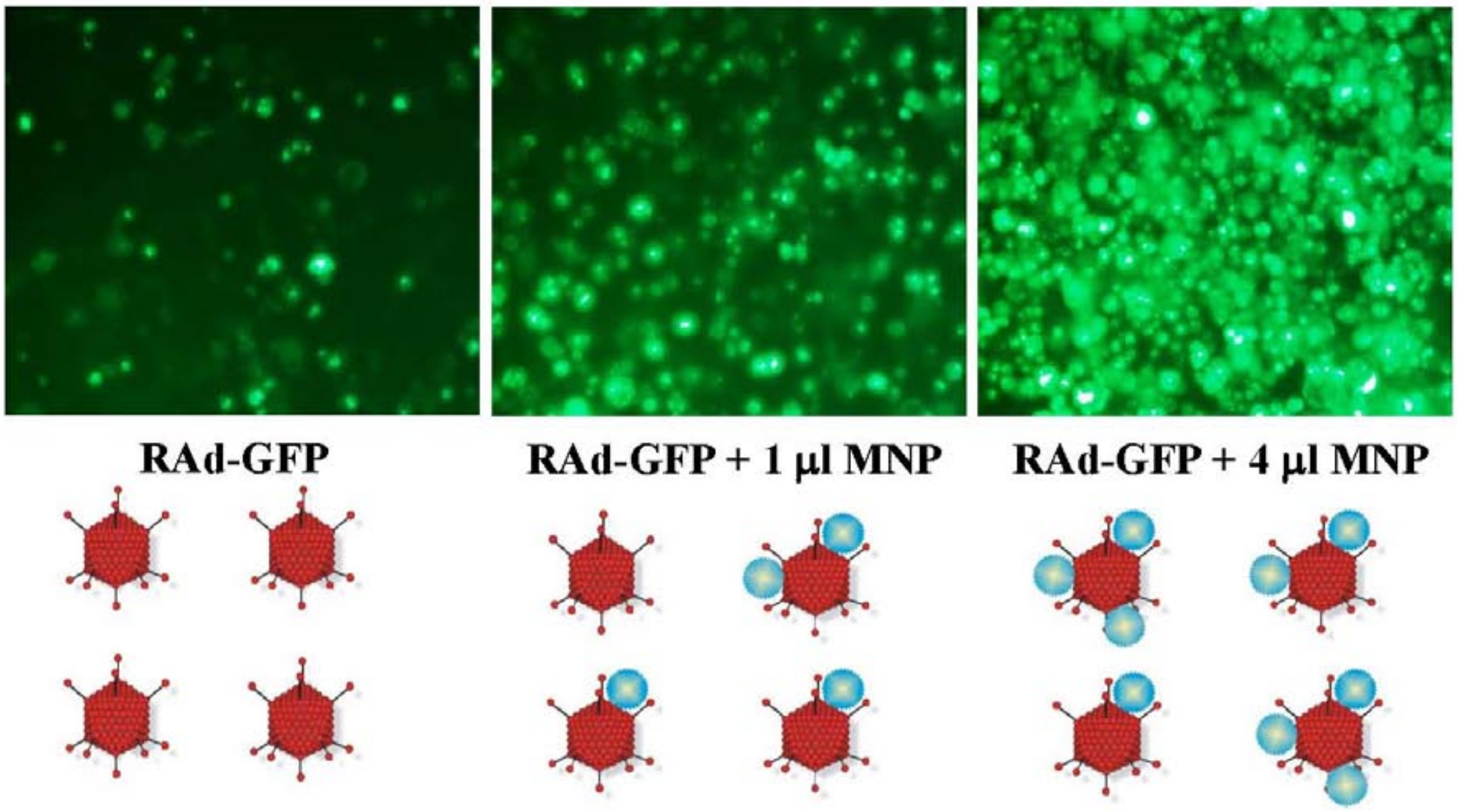

**Figure 3.-** Numerical simulation of magnetic field amplitude and field gradient for a typical cylindrical permanent NdFeSm magnet of 2 cm diameter and 1 cm in height. Left panel: induction field B mapping. The polarization is chosen along x axis. B color values are shown on the inset scale. Right panel: Induction field profile as a function of distance to the surface, along x- and y-directions. Inset: values of the spatial derivative dB/dx (proportional to the magnetic force) along x- and y-directions taken from the same simulations.

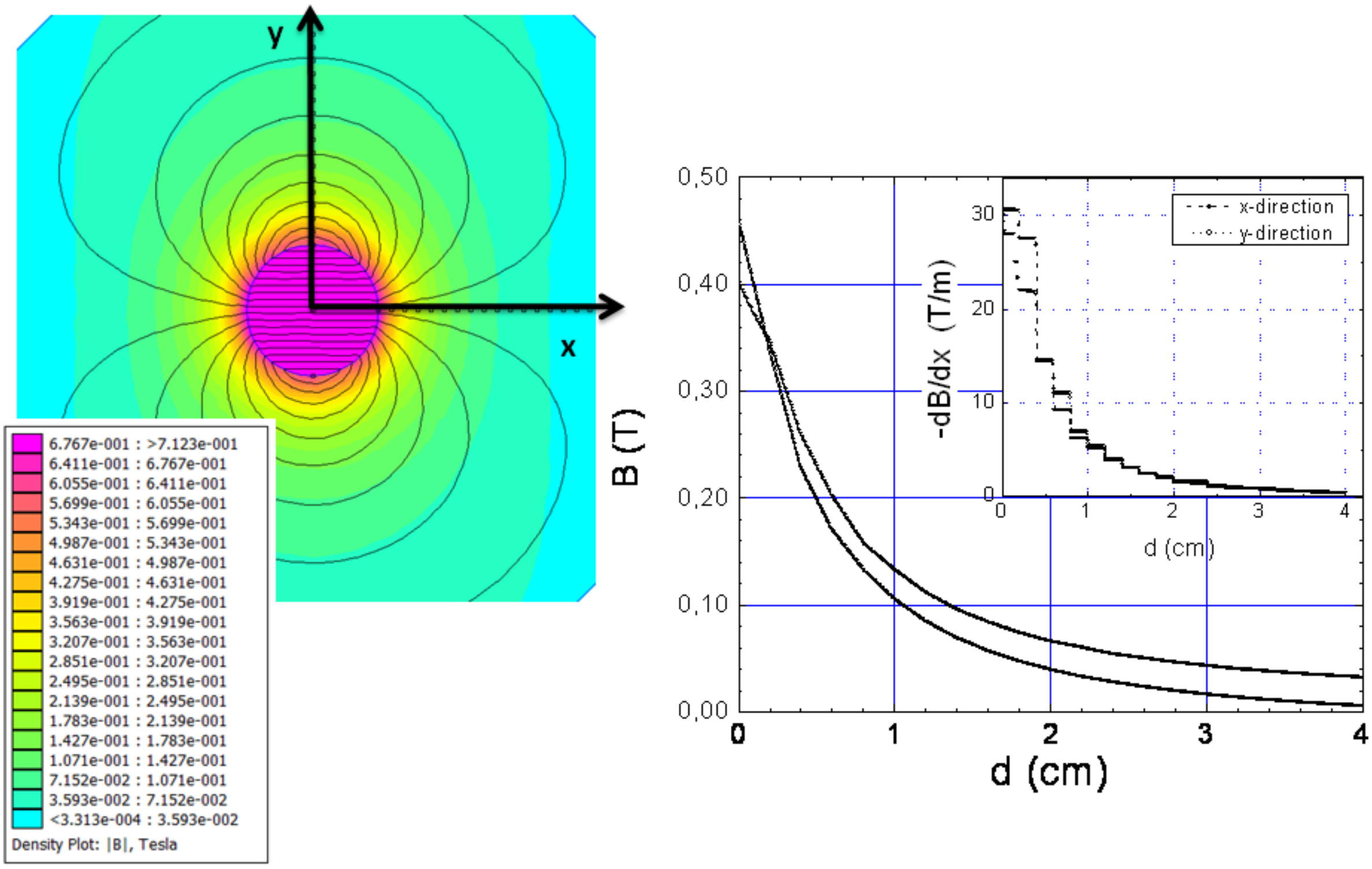

**Figure 4.-** Magnetite ($Fe_3O_4$) MNPs prepared by: (a) decomposition of $Fe(acac)_3$ in 1-octadecene, (b) precipitation-oxidation of $FeSO_4$ in aqueous media. **[Goya, GF et al., unpublished data]**

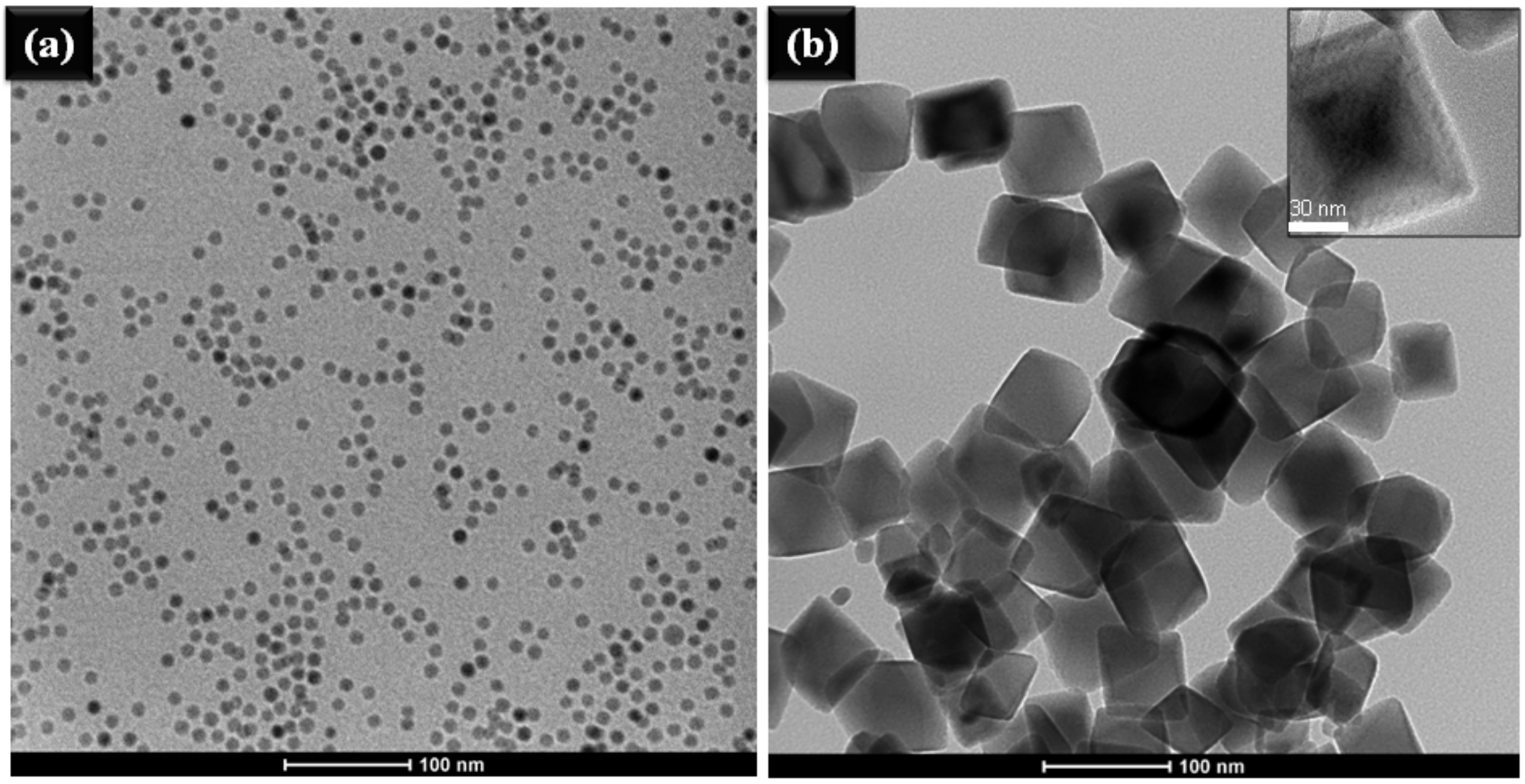

**Figure 5.- Proposed design for minimizing the invasiveness of gene therapy in the rat hypothalamus.** A MNP-RAd complex suspension is injected in the cisterna magna (5 µl) in the presence of a conical (or cylindrical) Nd-Fe-B permanent magnet placed in a proper orientation at the base of the rat head. The magnetic field drags the ferrofluid upstream the CSF flow towards the 3V (the target area) where the magnetic vector particles are concentrated so that the therapeutic transgene is delivered to the ependymal cell layer. After injection, the rat and the magnet are left in the same position for 30 min with the animal still under anesthesia. The magnetic field lines, magnet orientation and other details are intended for illustration only. They represent neither the precise configuration of the magnet nor the actual position for injection of head relative to the horizontal plane. The RAd virion is represented as a red icosahedron to which MNP (light blue spheres) are bound.
CM, cisterna magna; SA, Sylvian aqueduct; 3V, 3$^{rd}$ ventricle; LV, lateral ventricle **[Goya RG, *et al.*, unpublished]**.

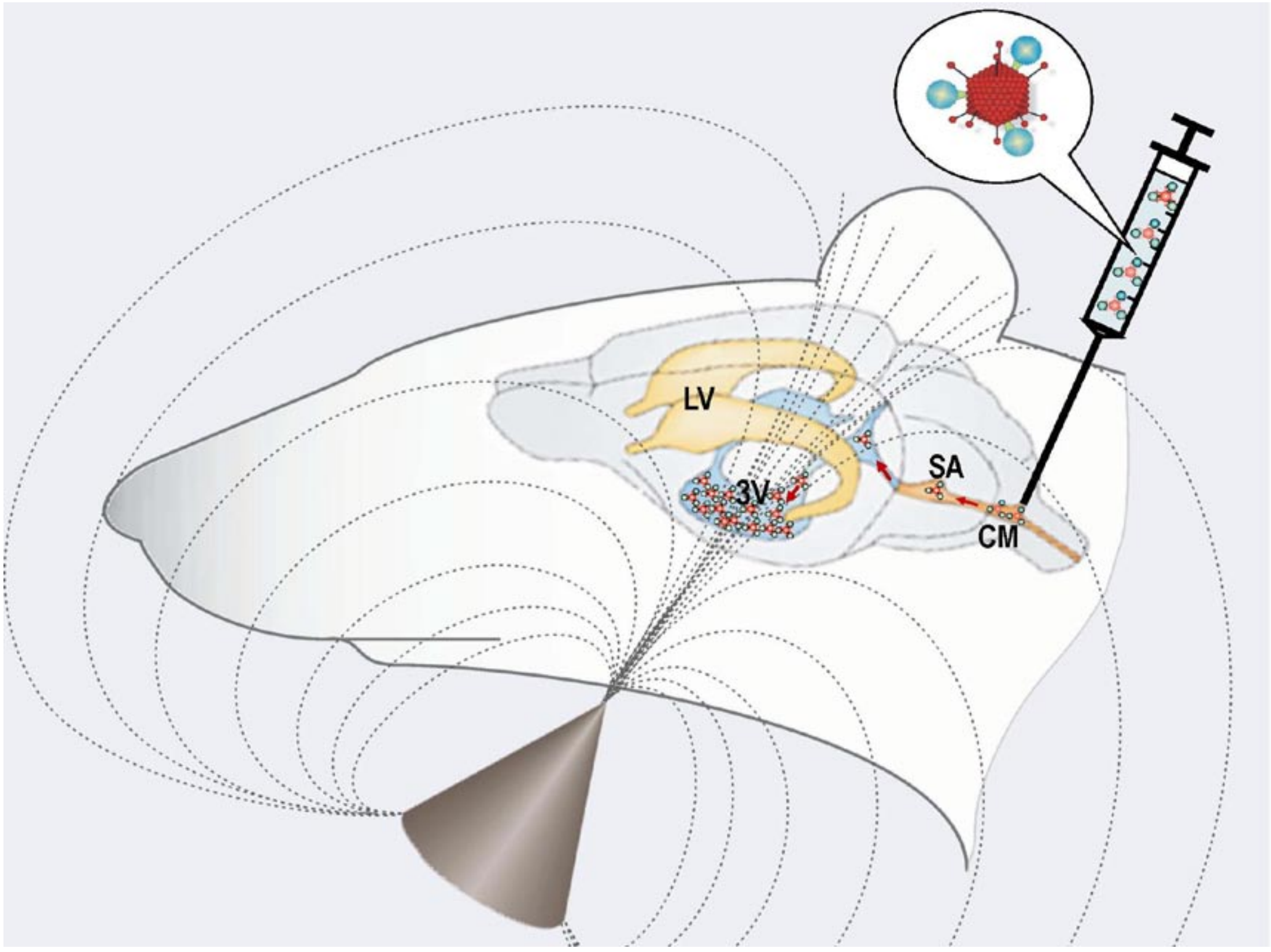

**Figure 6.- Advanced medical nanobots and current gene delivery nanoinstruments.-** ***Left panel:*** Artist´s view of future therapeutic nanobots injected into the blood stream [F**ront cover from ref. 98, with permission]** ***Right panel:*** Simplified diagram of a typical MNP-adenovector complex currently used for magnetofection.

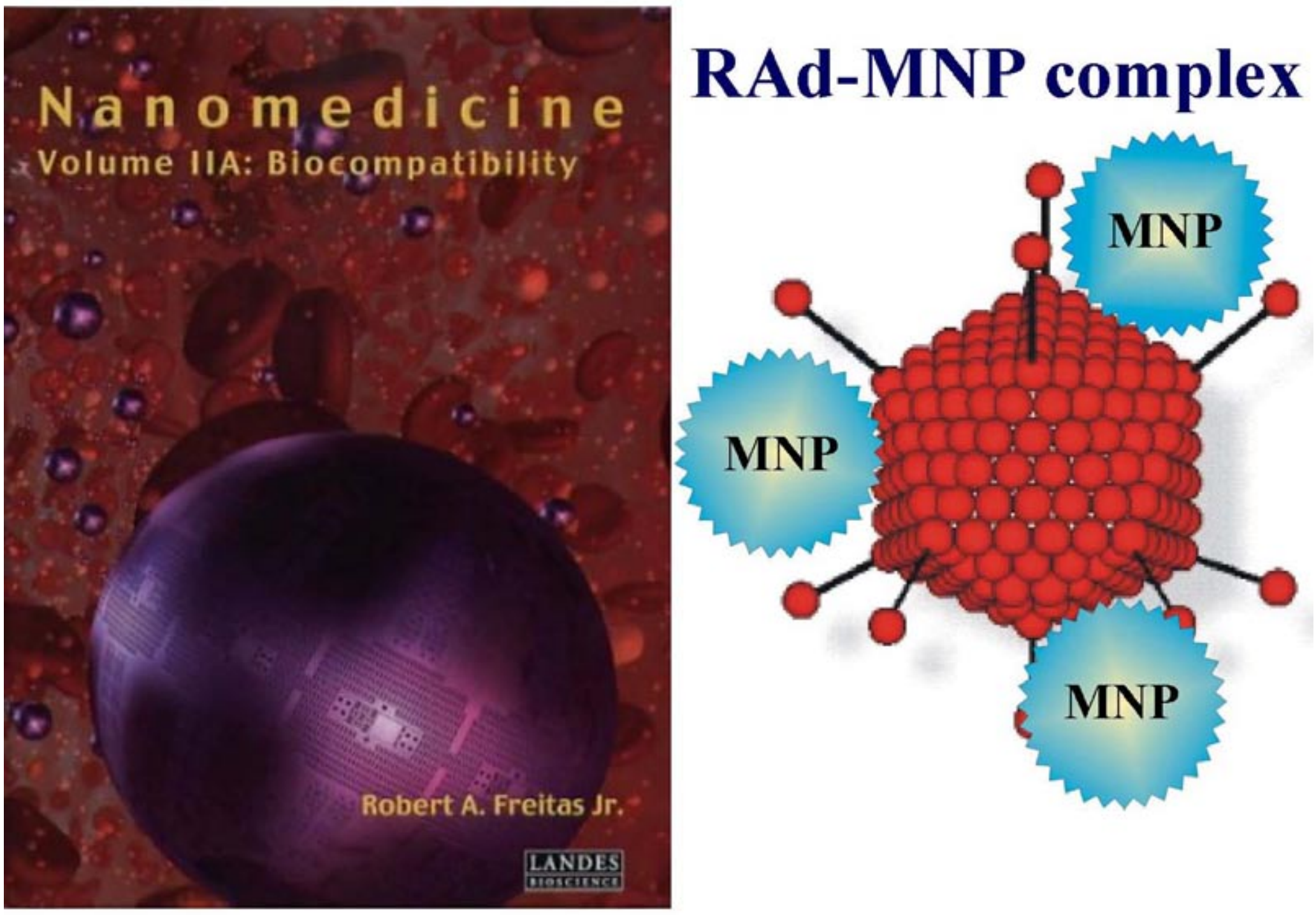